\documentclass[reprint, amsmath,amssymb,prper,showkeys]{revtex4-2} %linenumbers
\usepackage[utf8]{inputenc}
\usepackage[ngerman, USenglish]{babel}%for compiling problems in PRPER submission process

\usepackage{graphicx}% Include figure files
\usepackage{dcolumn}% Align table columns on decimal point
\usepackage{bm}% bold math
\usepackage{docs}%
\usepackage{bm}%

\usepackage[normalem]{ulem}
\usepackage[table]{xcolor}
\usepackage{dcolumn}
\usepackage{enumitem}
\usepackage[colorlinks=true,linkcolor=blue]{hyperref}

\begin{document}
\title{The Future Quantum Workforce: Competences, Requirements and Forecasts} 
\author{Franziska Greinert}
\email{f.greinert@tu-braunschweig.de}
\author{Rainer Müller}
\affiliation{Technische Universität Braunschweig, Institut für Fachdidaktik der Naturwissenschaften, Bienroder Weg 82, 38106 Braunschweig, Germany}
\author{Philipp Bitzenbauer}
\affiliation{Friedrich-Alexander-Universität Erlangen-Nürnberg, Professur für Didaktik der Physik, Staudtstr.~7, 91058 Erlangen, Germany}
\author{Malte S. Ubben}
\affiliation{WWU Münster, Institut für Didaktik der Physik, Wilhelm-Klemm-Str.~10, 48149 Münster, Germany}
\author{Kim-Alessandro Weber}
\affiliation{Leibniz Universität Hannover, Institut für Quantenoptik, Welfengarten~1, 30167 Hannover, Germany}

\date{\today}
\begin{abstract}
With the increasing industrial relevance of new quantum technologies, a well educated quantum workforce becomes increasingly crucial. The foreseeable lack of workforce raises important questions. 
What are the expectations regarding the future relevance of second generation quantum technologies? 
What are the requirements for the workforce in the coming quantum industry? Which competences, knowledge and skills should the future employees have?
In this paper, we report the results of our study that was aimed at mapping requirements and forecasts for the future quantum workforce. Our study consisted of three consecutive survey rounds. In total, we gathered $188$ responses from industry and academic experts across Europe. 
Our study results served as an input for the development of the European Competence Framework for Quantum Technologies, delivered by the project QTEdu CSA for the European Quantum Flagship. 
In addition, we will discuss predictions from experts related to the future quantum workforce, including the expected industrial relevance of the main areas of quantum technologies, the need for educational efforts, and the expected influence of quantum technologies on everyday life. 

\end{abstract}
\maketitle

\section{Introduction}
Quantum technologies are on the rise. Interest in the new 
quantum technologies (QTs) is growing and they are becoming relevant in industry~\cite{acin_quantum_2018}. 
Main QT areas are: (1)~Quantum communication, where single quantum objects are used to exchange information in a physically secure way~\cite{gisin_quantum_2007}. (2)~Quantum sensors, which use effects such as the behavior of a single quantum object in a magnetic field to make high-precision measurements~\cite{degen_quantum_2017}. (3)~Quantum computers, which are expected to have big impact if they reach a critical number of logical quantum bits (qubits)~\cite{gill_quantum_2022}. They promise the ability to solve some problems much faster than any classical computer, such as optimization problems~\cite{farhi_quantum_2014}, machine learning~\cite{ablayev_quantum_2020} or the factorization of large numbers, a problem on which the security of current cryptographic algorithms is based~\cite{shor_polynomial-time_1999}. (4)~Quantum simulations -- sometimes treated as a special part of quantum computation -- promise, for example, the ability to simulate large molecules, thus advancing quantum chemistry~\cite{daley_practical_2022}. 

The rapidly evolving field of these modern QTs is now in the phase of transition from a research topic to an industry-ready technology. 
This poses new challenges for the new emerging workforce who will develop  these QTs or work with it~\cite{kaur_defining_2022}. A considerable need for these experts is expected in the coming years~\cite{venegasgomez_quantum_2020} and now it is the time to start training this future quantum literate workforce~\cite{aiello_achieving_2021}.

In Europe, these efforts are driven by the Quantum Flagship~\cite{quantum_flagship_qteu_2022} and the corresponding Coordination and Support Action for Quantum Technology Education (QTEdu CSA)~\cite{qtedu_csa_qtedueu_2022} respectively the follow-up project Quantum Flagship Coordination AcTion and Support (QUCATS)~\cite{european_commission_quantum_2023}.
Likewise, in the US there are activities to elaborate the needs of industry~\cite{fox_preparing_2020, aiello_achieving_2021, hughes_assessing_2022, plunkett_survey_2020} and a new workforce development plan is initiated~\cite{subcommittee_on_quantum_information_science_qist_2022}. 

In the present paper, we report on our study that was conducted from March 2020 to May 2021. The main research objective was to collect and identify competences in the field of QTs, thus laying the foundation to compile the \textit{European Competence Framework for Quantum Technologies}~\cite{greinert_european_2021} in the QTEdu CSA. 

The Competence Framework aims to map all possible competences, knowledge and skills, that could be relevant for the future quantum workforce and its development. 
Meanwhile, it has already been used successfully, e.g. in the development of Qualification Profiles~\cite{greinert_qualification_2022} or in the preparation of EU-founded projects. Here the Competence Framework serves as a common language to map and compare modules and courses. 

In the follow-up project %of the QTEdu CSA, 
QUCATS%~\cite{european_commission_quantum_2023}
, the Competence Framework has been updated and extended by proficiency levels (version~2.0 from April 2023, for the latest version see doi \href{https://doi.org/10.5281/zenodo.6834598}{10.5281/zenodo.6834598}) and will be used as the basis for a certification scheme for QT training. Accordingly, our study provides important input for the planning of curricula and training programs on QTs.

~\\
In addition, we aimed to gather predictions about the future role of QTs and QT education, such as the relevance of QT industry training. Hence, we have two main goals, the following \textit{research objectives}:
\newcommand{\Rone}{Collection of requirements for the future quantum workforce, identification of domains and categorization of competences to prepare for the development of the European Competence Framework for QTs}
\newcommand{\Rtwo}{Derivation of predictions around the future industrial relevance and societal and educational impact of QTs}
\begin{itemize}[leftmargin=0.8cm] 
    \item[R1]  \Rone.
    \item[R2]  \Rtwo.
\end{itemize}
A general overview of our study and the methodology is provided in Sec.~\ref{sec:methods}, including the sample of our study. We continue with Sec.~\ref{sec:R2-competences} and~\ref{sec:R3-predictions}, which address the two research objectives, and close with a discussion in Sec.~\ref{sec:discussion}. The Appendix provides additional information on (\ref{app:expert})~sample  and (\ref{app:tabs}) results, as well as (\ref{app:CFoverv}) overview pages of the beta and 1.0 versions of the Framework.

\section{Methods\label{sec:methods}}
\subsection{Study design}
With our study we primarily  aimed on collecting expert opinions, input for the competence framework and statements around the future quantum workforce.
During the preparations of the study in 2019, the field of QT education for industry was in its infancy. Due to its broadness and complexity, no clear educational experts on this topic could be named, neither did specific literature exist for an initial framework. Therefore, we decided on an exploratory approach with consecutive questionnaires to the community, thus with an open (not pre-selected) expert panel. 

Invitations were sent to personal contacts and lists of people who had already shown interest in the topic, as well as -- increasingly as the study progressed -- to the growing Quantum Flagship communities, the Quantum Community Network (QCN) and the list of QT stakeholders (both were publicly available at Flagship website~\cite{quantum_flagship_qteu_2022}), and the QTEdu community~\cite{qtedu_csa_qtedueu_2022}. In addition, the European Quantum Industry Consortium (QuIC)~\cite{european_quantum_industry_consortium_euroquicorg_2022} distributed invitations to the main rounds to its members. Social media channels and the Quantum Flagship newsletter were also used to announce the study.  

Our iterative survey consisted of a small pilot round and two larger main rounds to successively open up the field:
\begin{itemize}[leftmargin=1.8cm]
    \item[pilot:]  overview, mainly open-ended questions, %(N$=28$)
    \item[main 1:]  refinement, e.g. pre-structured questions, %(N$=66$)
    \item[main 2:]  assessment, ratings of previous results. %(N$=94$)
\end{itemize}

We collected an initial overview in a pilot round (28 participants) that was carried out around March 2020. The questionnaire consisted of mainly open-ended questions to get an overview of the field. The result was the basis for the first main round (main~1) questionnaire. The main~1 questionnaire aimed to collect more concrete competences for the framework, based on the input from the pilot, and more community views on the future of and the related workforce.  In the two main rounds, the questions became increasingly structured and the focus shifted to the rating of scale items. The first main round took place in autumn 2020 with 66 participants, the second main round in spring 2021 with 94 participants. Within about one year, we thus collected a total of 188 responses.

The iterative approach has some similarities with the Delphi method~\cite{clayton_delphi_1997, hader_delphi-befragungen_2009}, as the subsequent questionnaire is based on the results of the previous questionnaire round. In its different types, it is an established method for different research aims like gathering a consensus or collecting expert opinions -- not necessarily with the aim of reaching consensus. The method has already been used successfully in many different areas of physics education research~\cite{hausler_physikalische_1980, weber_quantenoptik_2018, krijtenburg-lewerissa_key_2019}. 
Likewise, using a Delphi approach is a common practice for creating a competence framework~\cite{dijkman_competences_2017, dijkstra_developing_2021, muniz-rodriguez_developing_2017}. In empirically well-established fields a draft can be precompiled from literature research and is  validated in the Delphi study.

As a consequence of our approach, our study does not have a consistent experts panel, as it would be usual in a Delphi study, i.e.~the same group of people answering all questionnaires. 
Instead, we had a growing number of participants, although the majority only took part in one survey round each, the quantity of participants gives a broader picture of opinion.

The monitoring panel of our study consisted of the authors of this paper: a professor of physics education with more than 20~years of experience in quantum education (RM), three physics education researchers who already have PhDs in quantum education (PB, MU, KW), and a PhD student doing research on the development of the Competence Framework (FG).
Our task was to design the questionnaires and guide the process of collecting expert opinions by creating the next stage questionnaires founded on the previous results. For example, we had to make a selection of the statements from main~1 and decide which ones to provide for a rating in main~2 (details in Sec.~\ref{sec:R3-predictions}). Therefore, the process is not comparable to a completely free discussion of experts. 

\subsection{Instruments}
For our study, we used three online questionnaires created with the survey tool LimeSurvey~\cite{limesurvey_gmbh_limesurveyorg_2022}. The questionnaires were divided into separate parts. The first part addressed the participants' professional background, the other parts focussed on the two reasearch objectives. 

In addition to the initial qualitative data on competences and predictions gathered trough open-ended questions, we collected some quantitative data, primarily  from six-level rating scale items. For example, we asked to rate one's own competence, the future importance of QT in industry or the agreement to a statement from 1 `very high~/ very important~/ total agreement' to 6 `very low~/ totally unimportant~/ total disagreement'. In addition, there was a ranking question and single or multiple choices. For example, participants had to select the areas their profession covers.

In the pilot round, we focused on qualitative data collection. We used word clouds in order to provide a first insight into our data (shown in~\cite{gerke_quantum_2020}) and then conducted a qualitative content analysis~\cite{mayring_qualitative_2015} documented in~\cite{gerke_requirements_2022}. We also categorized some qualitative data in the two main rounds, always discussing the categorizations in the monitoring team. 
Additionally, we selected some (in our point of view) interesting comments and assessments. We showed them in the next questionnaire and asked in detail what the group of participants thought about them. 
In this selection process the influence of us, the monitoring team, is clearly visible. 

In main~1, we collected qualitative data mainly for R1 and thus input for the Competence Framework development in the QTEdu CSA~\cite{qtedu_csa_qtedueu_2022}. Here, we made the categorization for the content analysis together in the monitoring panel and discussed critical points to make a common categorization, as described in detail in Sec.~\ref{sec:R2-competences}. 

During data analysis, we connected for some questions the given answers with the ones on the professional background of a participant. For example, we compared the own competence rating with the agreement on a statement that a specific QT area will become the most important one, see Sec.~\ref{sec:R3-predictions} on R2.

\subsection{Expert panel \label{sec:participants}}
As the surveys were not restricted to a pre-selected expert group, we posed additional questions about the participants' background. With these, we aimed to better understand the experts panel. For example, we asked... 
\begin{itemize}[leftmargin=0.8cm] 
    \item[...] how the experts rate their own competence, or
    \item[...] from which area (industry, science or R\&D) they come. 
\end{itemize} 

These questions where the same for each round to keep them comparable.  
Details on the data collected on the expert panel can be found in the appendix \ref{app:expert}.

We decided for an open expert panel to gain as many opinions as possible on the new, rapidly developing field of QTs.
In the pilot, most participants had a scientific background, had many years of experience, and rated their competence (very) high. 
Between the pilot and the two main rounds we had a shift towards more participants from industry, more newcomers with not so many years of experience, but with insights what is important now -- and what will become important in the near future. Still, there are more participants from science than from industry, and more with theoretical knowledge than with practical/experimental skills. To get clearer insights into industrial needs,  future research needs to gather the opinions of more participants with a strong industrial background, e.g. in interview studies. 

\section{Collection of required competences\label{sec:R2-competences}}
\subsection{Methodology}
According to research objective R1, the main goal of our study was the collection of desired competences for the future quantum workforce as an input for the Competence Framework. For a first exploration, the pilot round questionnaire consisted of open-ended questions on which competences a future employee working with QTs will need. 
A qualitative content analysis inductively provided four central categories: (1) many answers relating to phenomena or basic principles, (2) fewer on mathematics, (3) some on physics and (4) a few on specific applications. A map showing a collection of sample answers from the pilot for these four categories (see~\cite{gerke_requirements_2022}) was given as an inspiration in the main~1 questionnaire, with the aim of inspiring the participants to be more specific, especially on the application aspects.

We adapted an item format from~\citet{hausler_physikalische_1980} comprising three aspects of the competences: To derive as concretely defined competences as possible, our participants were asked to first decide on a specific subfield in which they had expertise. For this specific subfield, the participants had to... 
\begin{itemize}[leftmargin=0.8cm]
    \item[...] formulate a concrete competence, 
    \item[...] describe what this competence is useful for, and 
    \item[...] determine the level of expertise required for users~(U) or developers (D), respectively.
\end{itemize}

The participants got an example in the form of a table, which is reproduced in Tab.~\ref{tab:exComp}.
\begin{table}[htb]
    \centering
    \begin{ruledtabular}
    \caption{Example of a competence in the three-aspect structure from the main~1 questionnaire for the subfield \textit{software development}.}
    \label{tab:exComp}
    \begin{tabular}{p{2.05cm} p{1.95cm} p{4cm}}
    competence & useful for & needed level of expertise \\
    \hline
    understanding of qubit operations and quantum gates & composing quantum algorithms and applying them to specific tasks 
    & \textbf{U:} deeper basic knowledge of the qubit concept and the effects of different operators on a formal-logical level. No specific knowledge of physical implementation of the operators and the qubits themselves is needed.
    \end{tabular}
    \end{ruledtabular}
    %\vspace{-6pt}
\end{table}

\begin{figure}[hbt]
    \centering
    \includegraphics[width=\linewidth]{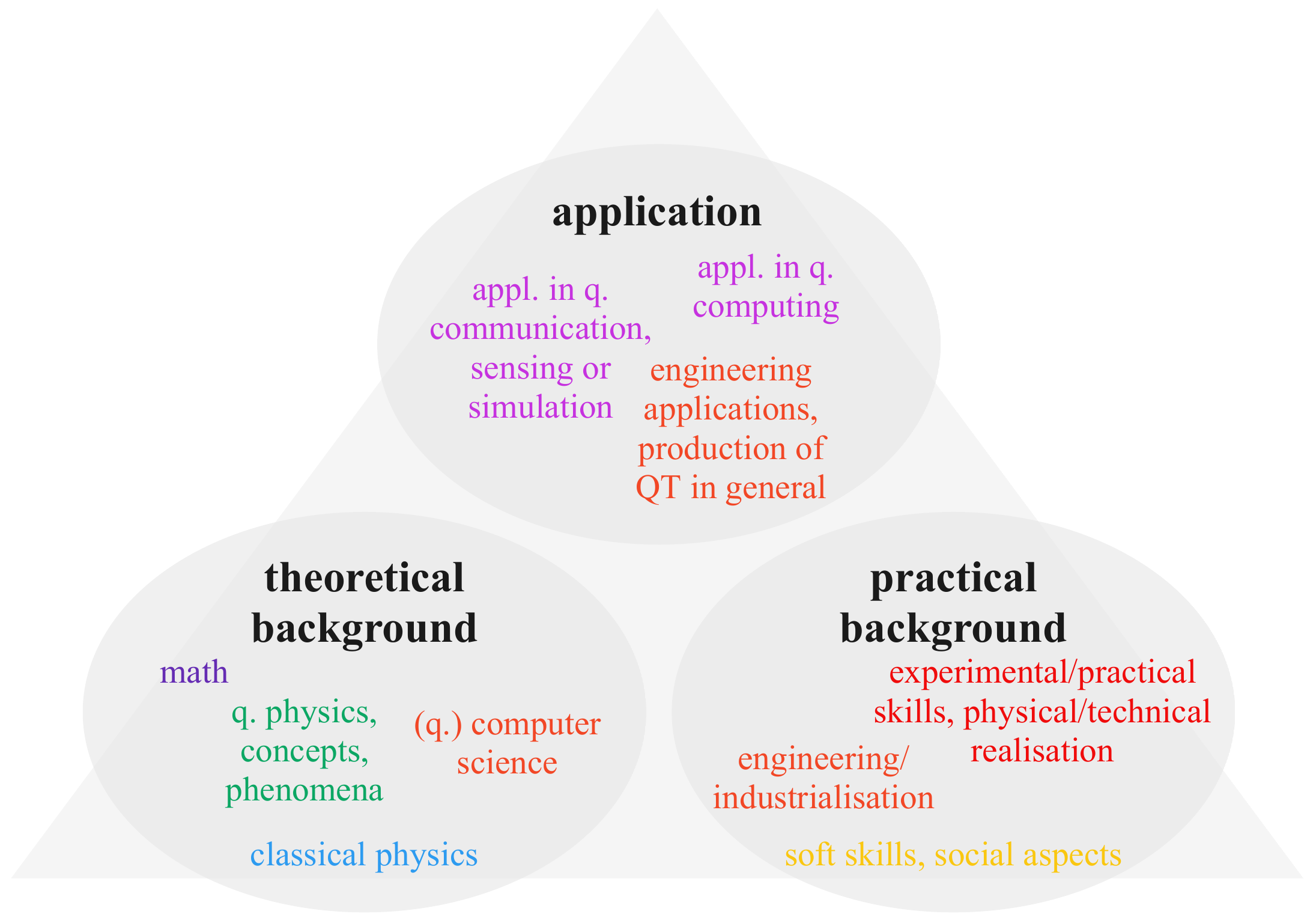}
    \caption{Three main categories from the qualitative content analysis in main 1 with associated subcategories. The colouring of the subcategories corresponds to the one used during the categorization process. Same colours indicate subcategories that were separated later in the process, e.g.~the splitting of (quantum) computer science from engineering.}
    \label{fig:categories}
\end{figure}

In this way, we collected 183 responses for 56~subfields (i.e., 56~people answered this question) and categorized them through qualitative content analysis. Based on the data set, we inductively derived a category system consisting of three main categories: the theoretical background including subcategories of classical and quantum physics, mathematics and computer science; the practical background with experimental skills, engineering and soft skills; and the applications where the QTs come in, as visualized in Fig.~\ref{fig:categories}. The category system with descriptions of the subcategories and examples is shown in Tab.~\ref{tab:categoRulesExamples} in the appendix~\ref{app:tabs}.

\subsection{Results}
The categorization statistics of the 183 responses collected in main~1 are shown in Tab.~\ref{tab:catego}. As several responses are assigned to more than one (sub-)category, especially a theoretical or practical background and an application subcategory, the total number sums up to a higher number than 183.

\begin{table}[ht]
\caption{\label{tab:catego} Categorization statistics from the qualitative content analysis in main~1 with the number of responses categorized in the corresponding subcategory.}
\begin{ruledtabular}
\begin{tabular}{lp{6.3cm} d}
& subcategory & \text{N}\\
\hline
\multicolumn{3}{l}{\textit{theoretical background}}\\
&quantum physics, basic concepts, phenomena&110\\
&classical physics&24\\
&mathematics&33\\
&(quantum) computer science&45\\
\multicolumn{3}{l}{\textit{practical background}}\\
&experimental/practical skills, physical/tech\-ni\-cal realisation&43\\
&engineering/industrialisation&16\\
&soft skills, social aspects&22\\
\multicolumn{3}{l}{\textit{application (`useful for' part)}}\\
&engineering applications, production of QT in general&68\\
&application in quantum computing&60\\
&application in quantum communication, sensing or simulation&25\\
\multicolumn{2}{l}{\textit{other}}&2\\
\end{tabular}
\end{ruledtabular}
\end{table}

\subsection{Discussion}
The categorization statistics for the competences from main~1 (Tab.~\ref{tab:catego}) document many responses on quantum phenomena and the basic concepts of quantum physics. They were categorized together with the few answers on traditional quantum physics in the first subcategory with by far the most entries. For mathematics and classical physics there were far fewer answers, but still so many that their relevance is clearly visible. 
In comparison to the pilot round results, we note the occurrence of (quantum) computer science with even more answers than on mathematics, for example. Many answers included concrete applications, which is what we wanted to achieve with the `useful for' part of the question. 

Regarding the applications, a strong focus on quantum computing can be observed. There were more than twice as many answers on this QT area than on the other three main QT areas communication, sensing/metrology and simulation combined.
In addition, practical and soft skills, including engineering aspects, showed up as essential for the future quantum workforce. Thus, we found a three-part structure: QTs/applications, theoretical background, and practical background, where applications are central. This is visible in Fig.~\ref{fig:categories}.

\begin{figure}[ht]
\centering
\includegraphics[width=\linewidth]{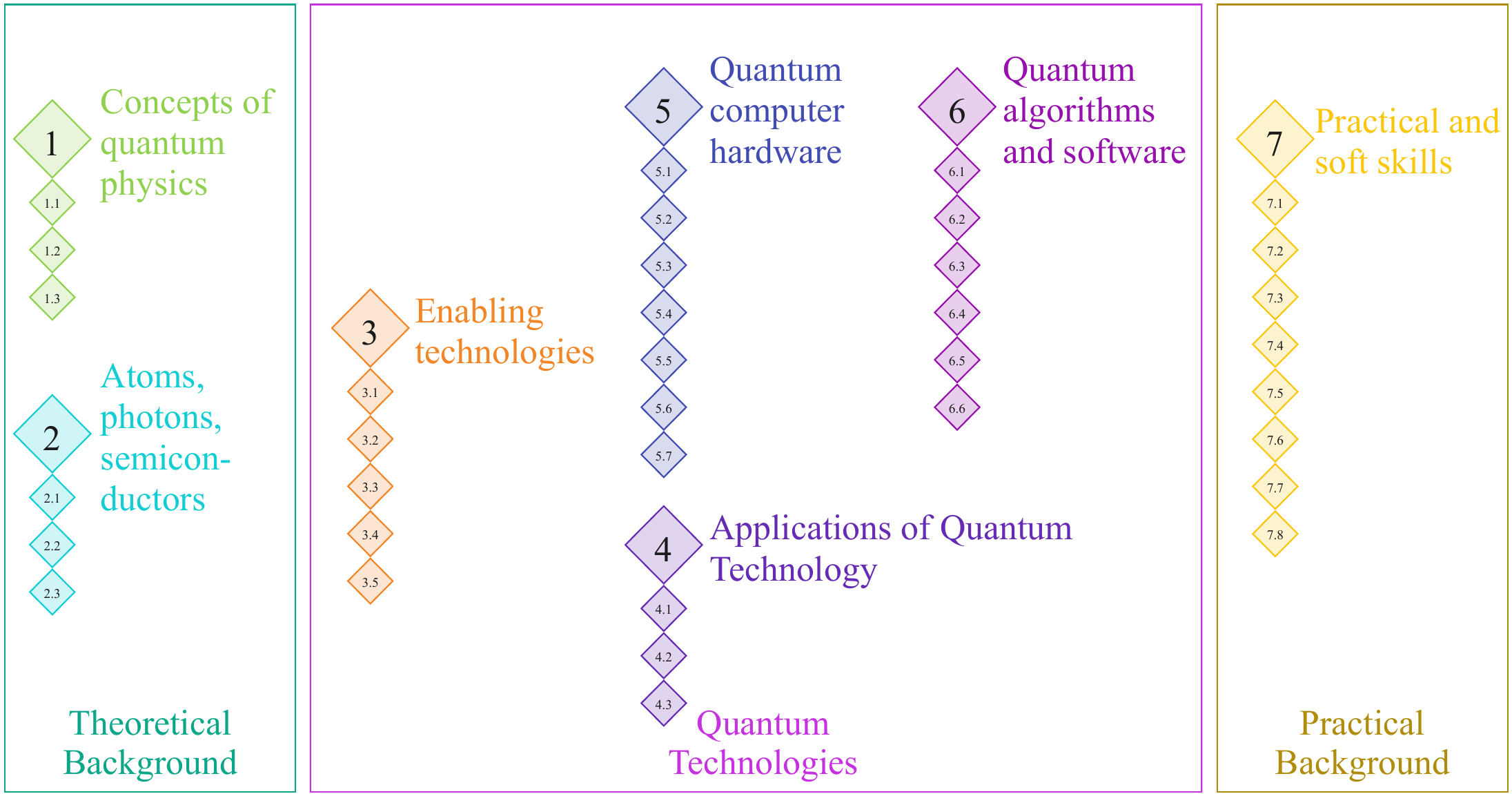}
\caption{Structure of the beta version of the European Competence Framework for Quantum Technologies~\cite{greinert_nee_gerke_beta_2020}. For reasons of readability, only the seven main domains are explicitly shown. The anchor points (1.1, 1.2, etc.) for the subdomains give an impression of the overall structure (full-page version with readable subdomains in the appendix, Fig.~\ref{fig:CF-beta-full}.).\label{fig:CF-beta}}
\end{figure}

The categorization from main~1 was the starting point for the formulation of the European Competence Framework for Quantum Technologies. With an iterative sorting, structuring, recategorization and resorting process, the framework's beta version was compiled. The addition of more sub-points in the framework and sample statements from the answers of main~1 completed the beta version, which was published in December 2020~\cite{greinert_nee_gerke_beta_2020}. 
Fig.~\ref{fig:CF-beta} shows  
the basic structure of the beta version.  More details on the development can be found in~\cite{greinert_competence_2021, gerke_vh_greinert_ermittlung_2021}.

The three-part structure from the categorization (Fig.~\ref{fig:categories}) with the QT applications central, flanked by the theoretical and the practical background is clearly visible. Also, some of the colors used in the categorization remained in the framework's beta version. This shows the emergence of the framework's structure from the categorization of the responses collected in our study.

The beta version reflects the categorization statistics from main~1: it had dedicated domains for computing hardware (domain 5) and software (6), but only one combined domain for the other QTs communication, sensing and simulation (4). The focus on computing is obvious.

The beta version was shared in the community for feedback, especially in the QTEdu working group kick-off meetings in March 2021~\cite{greinert_competence_2021, qtedu_csa_qtedueu_2022}. The over-representation of computing was criticized and led to structural changes in the application/QT block, to dedicated domains for sensing and communication, inclusion of simulation in the computing software domain and a combined hardware domain for computing and sensing~\cite{greinert_competence_2021}. More details on the further framework development in addition to the presented study will be discussed at the end of this paper.

\newcommand{\RQ}[1]{\begin{itemize}[leftmargin=0.8cm]
    \item[$\blacktriangleright$] #1
\end{itemize}}

\section{Predictions\label{sec:R3-predictions}}
\subsection{Methodology}
In order to approach research objective R2, we collected input in the pilot round as well as in main~1. We mainly used rating scale items on the \RQ{future relevance of the main QT areas  in certain time periods,} but also open-ended questions where participants had to argue \RQ{which technology will become important for what reason.}
Based on this input, we designed rating scale items for the main~2 questionnaire. 

In the four pillar structure of the Quantum Flagship~\cite{quantum_flagship_qteu_2022, acin_quantum_2018}, modern QTs are divided into communication, computation, simulation and sensing/metrology. We followed this structure in formulating the questions.  However, in recent descriptions of QTs and also in the Competence Framework version~1.0~\cite{greinert_european_2021}, simulation is regarded as a part of computation.

From the pilot we concluded that QTs are already important, also in industry, but not as important as they will be in the next 5 to 10 years (from year 2020), as documented in~\cite{gerke_quantum_2020}. In addition, there were comments assessing quantum computing as the most important QT, but in the rating for the industrial relevance of the four main QT areas, quantum computing was ranked lowest for the relevance in 5 to 10 years. This led to some more questions in main~1, where we asked for \RQ{the expected industrial importance of the four main QT areas and enabling/basic technologies in short (5 to 10 years) and in long term.} 
Again, quantum computing was ranked lowest in the short term, but close to the top in the long term, see~\cite{gerke_vh_greinert_ermittlung_2021}. 
This led to the inclusion of items in main round 2 questionnaire that address this issue: 
\textit{In the near future, quantum computing will be less important than the other QT areas} (M1) and \textit{In the long term, quantum computing will become the most important QT area} (M3). 
Based on the previous results, a total of 17 statements were formulated and given for a rating in main~2. The ratings were on a six-level scale from total agreement~(1) to total disagreement (6).

We assigned each item an identifier in order to provide comprehensive but well-arranged insights into our data: P1, P2, P3 for the statement based on data from the pilot, M1, M2, ... for the statements based on data from main~1 and Q1, Q2 for two quotes from main~1: 

\begin{itemize}[leftmargin=0.8cm]
    \item[P1] The relevance of QTs for industry will increase significantly in the near future.
    \item[P2] QTs are already very important in science, but even here they will become more important in the next few years.
    \item[P3] The relevance of QTs for society will increase significantly in the near future.
    \item[M1] In the near future, quantum computing will be less important than the other QT areas.
    \item[M2] In the near future, quantum simulation will be less important than the other QT areas.
    \item[M3] In the long term, quantum computing will become the most important QT area.
    \item[M4] In the long term, quantum sensing/metrology will become the most important QT area.
    \item[M5] In the long term, quantum communication will become the most important QT area.
    \item[M6] Quantum computation has the ``highest gain potential'' of all QT areas. In the long-term, it ``will have more impact and will really be disruptive.''
    \item[M7] Quantum simulation will have ``enormous long-term value for chemistry, pharmacy, material science, etc.''
    \item[M8] Quantum sensors/metrology will become very important through use in medicine (e.g.~imaging).
    \item[M9] Quantum sensors/metrology will become very important through use in timing/navigation, observation and autonomous devices/AI.
    \item[M10] Quantum communication will become very important because of cryptography/security and use in secure communication in banking, military, politics, etc.
    \item[M11] Quantum communication will become very important in the context of the quantum internet.
    \item[M12] Enabling/basic technologies will be the first to become really important in industry, as ``the industrial impact of QT can only been realized when quantum engineering, integration, miniaturization and scaling is realized'', so thier role is ``moving QT from the lab into society, making it aviable at reasonable cost''.
    \item[Q1] ``For me it is a question of maturity and opportunity window... all the disciplines will be important in the short term... those more matured and more deployed will lose their `importance' because they would have been absorbed, accepted and assimilated in the long term.... other will continue in the top in the long term due to their inmaturity or potential of evolution still for develop...''
    \item[Q2] ``In my opinion quantum communication including quantum internet will remain a merely academically interesting field of technology assuming that the only application which will be found for it is quantum-secure communication. The reason for this is that post-quantum-crypto systems (which are quantum-safe but classical alternatives to our existing crypto systems) will provide the solution for the risk which quantum poses to existing crypto-systems. Therefore, unless you have some national security type communication, quantum-key-distribution will always remain an unnecessarily expensive alternative to PQC systems. The other QT will in my opinion in the mid/long-term provide important contributions to business and society''
\end{itemize}

For each of these statements, besides the participants' agreement, we asked for the participants' response certainty, again a six-level scale from very sure to very unsure. Thus, we wanted to \RQ{gain deeper insights into the expected quality of expert predictions.}
For all questions, more than 60\% and up to 93\% of the participants assessed their voting as rather to very sure. Thus, the experts were sufficiently confident of their assessments.

Furthermore, in order to \RQ{include opinions and topics that we as a monitoring team had not foreseen and asked about, but to represent the field as broadly as possible,} 
we collected comments and remarks in the pilot round and provided them as items in a ranking question in main~1. In addition, we collected some more comments in main~1.
These led to 15~more statements S1, S2, ... to rate in the main~2 questionnaire.

Here, explicit reference is made to ``1st~gen QTs'', quantum technologies of the first generation that are based on effects to which multiple quantum objects contribute, such as lasers or semiconductor electronics, or ``2nd gen QTs'', modern technologies that make use of fundamental quantum effects such as superposition or entanglement of using single quantum objects~\cite{schleich_quantum_2016}. Since the focus of this paper is on 2nd generation QTs, we do not always state this explicitly in the text, instead we just use ``QTs''. This is common in public literature, where only ``QTs'' is usually used when reporting on new developments in 2nd generation QTs. However, since we also had statements about e.g. the emergence from 1st generation QTs to 2nd generation QTs, we explicitly refer to the QT generation in the statements S1 to S15:

\begin{itemize}[leftmargin=0.8cm]
    \item[S1] Quantum chemistry will be the most important subfield of 2nd gen QTs.
    \item[S2] The technological change of paradigm, i.e. the extension of current technologies to hybrid systems, is a really important aspect of 2nd gen QTs.
    \item[S3] Fundamental research becomes less important within 2nd gen QTs.
    \item[S4] 2nd gen QTs will enable further steps in fundamental research.
    \item[S5] It is more important to push 1st gen QTs to make 2nd gen emerge on this basis than pushing 2nd gen directly.
    \item[S6] In practice the emergence of 2nd gen from 1st gen is not essential.
    \item[S7] The interaction and integration of classical and quantum systems will be in focus of 2nd gen QTs.
    \item[S8] Decoherence is one of the most central challenges to be addressed in the realisation of 2nd gen QTs.
    \item[S9] It will be necessary to transform 2nd gen QTs from a research subject to a subject of everyday life.
    \item[S10] It will be necessary to communicate about the transformation of 2nd gen QTs from a research subject to a subject of everyday life (outreach).
    \item[S11] Creating networks between research groups and industry will be essential.
    \item[S12] Special educational programs fitted to arising needs are necessary.
    \item[S13] 2nd gen QTs will contribute to solve everyday problems.
    \item[S14] 2nd gen QTs will contribute to solve social challenges.
    \item[S15] 2nd gen QTs will lead to social inequality.
\end{itemize}

\begin{figure*}
    \centering
    \includegraphics[width=\linewidth]{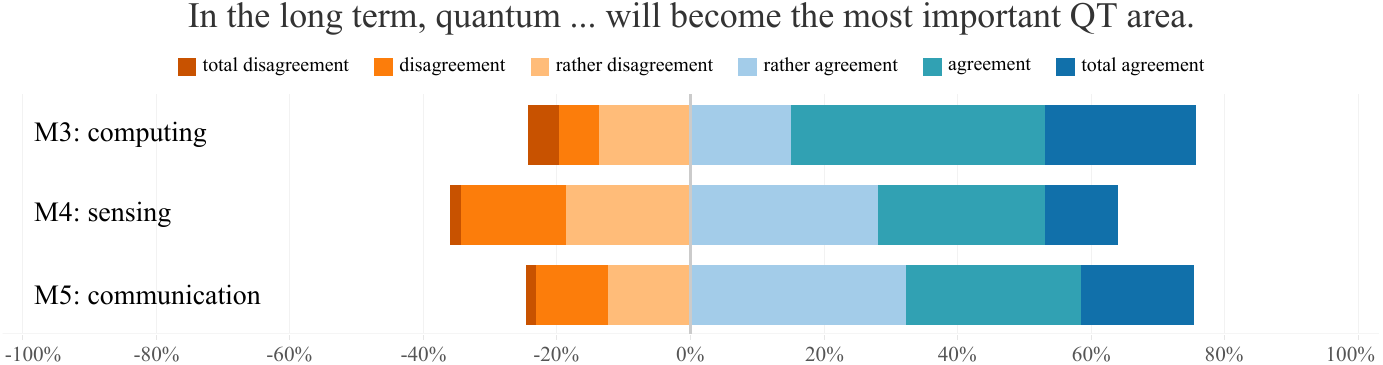}
    \caption{DSBC for the data from Tab.~\ref{tab:statem-under}. The part of the bars on the right side visualize the agreement, the ones on the left side show disagreement.}
    \label{fig:DSBC}
\end{figure*}

We used Diverging Stacked Bar Charts (DSBC)~\cite{robbins_plotting_2011} created with Tableau software~\cite{tableau_software_llc_tableaucom_2022} to present interim results on expectations of the  future relevance of the QT areas in an illustrative way, see~\cite{gerke_quantum_2020, gerke_vh_greinert_ermittlung_2021}. 
In these charts, all votes on rather to total disagreement (rating 4, 5 or 6) are located on the left  side of the midline, and all votes for rather to total agreement (rating 3, 2 or 1) are  on the right  side.
The shift between the bars, and thus between the agreement ratings, is clearly visible. Here, a DSBC is used to visualize the expectations of which QT will become the most important one in the long run, see Fig.~\ref{fig:DSBC}.

The different assessments of short and long term relevance on the future QT areas led to another research question addressed in the main~2 questionnaire: \RQ{Do the experts prefer a (nearly) equally distributed educational effort on all QT areas or would they set a strong focus -- and which one?} 
We asked: \textit{What proportion of quantum education do you think should be allocated to the following application/context areas: quantum computing, sensing, communication or simulation? Enter the percentage for each (without `\%').}
The answer layout were four fields, one for each area, were only numbers could be entered, and with a sum control so that 100 could not be exceeded. 

For showing the answers in a compressed way, we decided to cluster them in three groups. As most answers were at 20\% to 30\%, and this is also the area where nearly equally distributed answers are located, they form the central group. The other answers on more than 30\% or less than 20\% form the other two groups. So the answered numbers are clustered in `$>30$', `$20-30$' and `$<20$' (in percentage).

\subsection{Results}
The rating statistics of the responses for all 32 statements from the main~2 questionnaire are listed in Tab.~\ref{tab:statem-over} and Tab.~\ref{tab:statem-under} in the appendix~\ref{app:tabs}. The order of the items is determined by the proportion of agreement, i.e.~which percentage of the participants agreed rather to totally with the statement (rating 3, 2 or 1).

The highest level of agreement was obtained for P1, which predicts that \textit{the relevance of QTs for industry will increase significantly}, followed by S12 on the necessity of \textit{special educational programs}.

Tab.~\ref{tab:statem-over} contains the 18 statements that gained more than 80\% agreement. They all have one clear maximum in the number of votes: a rating of 1 (total agreement) or 2 (agreement) that was selected by more than one third of the participants. These maxima are marked in the table in the appendix. Also the first eight of the other 14 statements (Tab.~\ref{tab:statem-under}) follow the same pattern showing a  maximum -- now on rating 2 or 3. All these statements obtained more than 50\% agreement. The last statement, S3 saying \textit{fundamental research becomes less important within 2nd gen QTs}, also has one maximum, but on rating 6 (total disagreement). 

The remaining statements M1, M2, S5, S6 and S15 received agreement from 40\% to 50\% of our study participants. They show two clusters, one on the agreement side with a maximum on rating~2 or~3 and one on the disagreement side with a maximum on rating~5 or~6. 

For the curriculum distribution, at least 60\% of the participants preferred to assign between 20\% and 30\% of the efforts on quantum sensing, communication and simulation. The others mainly voted for less than 20\% for these three areas.Only a few participants voted for a stronger focus in sensing or communication. For computing, only a few of the participants preferred less than 20\% of the efforts here, and the others split about half and half for 20\% to 30\% and more than 30\%. Details are shown in Tab.~\ref{tab:curriculum} that also lists mean and median of the assessed distributions. 

\begin{table}[hbt]
\vspace*{-\baselineskip}
\caption{\label{tab:curriculum} Preferred distribution (in percentage) of quantum education on applications of the four main QT areas. Numerical responses clustered to the groups of more than 30\%, 20\% to 30\% and less than 20\% of the educational efforts for the according area, with sensing including metrology and imaging, for $ \text{N}=67$ responses.}
\begin{ruledtabular}
\begin{tabular}{l|d@{\hspace{-.3cm}}d@{\hspace{-.5cm}}d@{\hspace{-.5cm}}|cc}
&> 30&20 - 30& <20&mean&median\\
\hline	
q computation&43&48&9&34&30\\ 		
q communication~~&15&64&21&24&25\\
q sensing&9&69&22&22&20\\	
q simulation&0&60&40&18&20\\
\end{tabular}
\end{ruledtabular}
\end{table}

\subsection{Discussion}
The statements cover a wide range of topics. Therefore, this discussion is divided into individual parts, for which only the results on the respective topic are discussed.

\subsubsection{What will be the most important quantum technology?}
Based on the answers from main~1, we stated: \textit{In the long term, quantum computing} (M3) / \textit{sensing/me\-trol\-o\-gy}~(M4) / \textit{communication} (M5) \textit{will become the most important QT area}. All three statements had agreement of more than 60\%, see Tab.~\ref{tab:statem-under}. Around 70\% of the participants agreed to more than one of these statements (rating 1, 2 or 3), and about 30\% were unsure about their assessments. This shows the significant ambiguity in this prediction. 

Fig.~\ref{fig:DSBC} visualizes the agreement distribution for these statements. 
The length of the bars on the agreement side are nearly equal for computing and communication, but the parts on agreement and total agreement are  larger for computing than for communication. Nonetheless, the part on total disagreement is larger for computing than for communication as well.

In addition, the statement that \textit{quantum computing has the `highest gain Potential' of all QT areas}~(M6), which was based on a comment from main~1, obtained higher agreement than the three statements discussed above, see Tab.~\ref{tab:statem-over}. This data set indicates quantum computing as most likely candidate to become the most important QT area, followed by quantum communication and then quantum sensing/metrology. 

However, the agreement to what  will be the most important QT might be biased by the participants' own expertise: A closer look on the participants´ expertise (see Tab.~\ref{tab:owncompM2}) reveals a similar picture as the agreement ranking. There is more high self-assessed expertise in computing (63\% on rating 1, 2 or 3) than in communication (53\%) or sensing/metrology (43\%). 
Thus we can not identify a definitive favourite for the long term most important QT. We can just conclude that likely the tendency is towards quantum computing or quantum communication, not quantum sensing/metrology.

\subsubsection{Future importance of the main QT areas}
\begin{figure}[hbt]
    \centering
    \includegraphics[width=\linewidth]{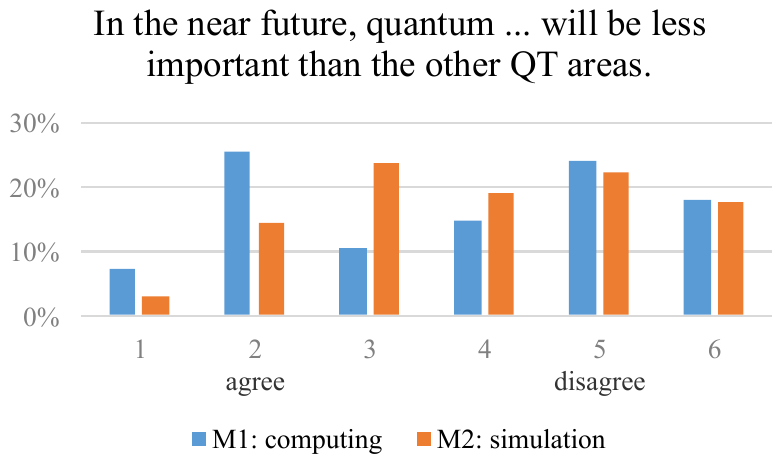}
    \caption{Distribution of agreement resp. disagreement to the statements M1 (for $\text{N}=67$ ratings) and M2 ($\text{N}=63$) for the data from Tab.~\ref{tab:statem-under}. }
    \label{fig:M1M2}
    % ~\\
    % \includegraphics[width=\linewidth]{Chart_S5S6.pdf}
    % \caption{Distribution of agreement resp. disagreement to the statements S5 (for $\text{N}=50$ ratings) and S6 ($\text{N}=48$) for the data from Tab.~\ref{tab:statem-under}.}
    % \label{fig:S5S6}
\end{figure}

For the \textit{near term relevance of quantum computing}~(M1) and \textit{quantum simulation} (M2), the ratings were quite heterogeneous, see Fig.~\ref{fig:M1M2}. 
Their relevance in the near future is therefore controversial for now.

The statements M7 to M12 are based on comments from main~1 on why the participants expect a particular QT area to become very important. For all of them the agreement was quite similar, about 40\% of the votes were on rating~2 (agreement) and the distribution of votes is alike, see Tab.~\ref{tab:statem-over} (Tab.~\ref{tab:statem-under} for M11). This indicates that these statements describe reasons for the future relevance.

\begin{figure}[hbt]
    \centering
    \includegraphics[width=\linewidth]{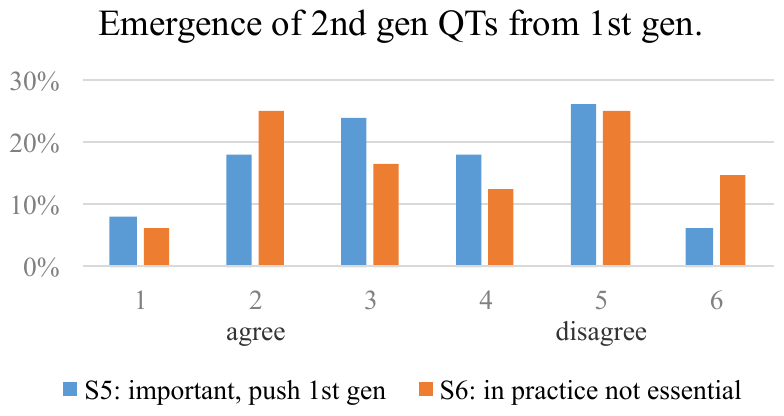}
    \caption{Distribution of agreement resp. disagreement to the statements S5 (for $\text{N}=50$ ratings) and S6 ($\text{N}=48$) for the data from Tab.~\ref{tab:statem-under}.}
    \label{fig:S5S6}
\end{figure}

In statements M12 the relevance of the enabling technologies received high agreement. However, whether \textit{it is more important to push 1st gen QTs to make 2nd gen emerge on this basis than pushing 2nd gen directly} (S5) or \textit{in practice the emergence of 2nd gen from 1st gen is not essential} (S6) is ambiguous, see Fig.~\ref{fig:S5S6}. 

For both statements, there are two clusters in the response statistics, one on the agreement side of the scale and one on the disagreement side, and around 50\% agreement for both. In addition, these two statements were the two with the smallest total numbers of assessments. This could be interpreted as a sign of uncertainty in the experts group. Here we see a very divided opinion among our experts.

\subsubsection{Educate the future quantum workforce}
Almost all participants agreed that \textit{the relevance of QTs for industry will increase significantly in the near future}~(P1), see Tab.~\ref{tab:statem-over}. This shows the relevance of getting industry workforce ready to work with quantum technologies. For the education of the future quantum workforce, \textit{special educational programs fitted to arising needs are necessary}~(S12). More than the half of the participants agreed totally on this need (rating~1), and only one person disagreed. These data clearly indicate an urgent need to train the future quantum workforce.

Not only special educational programs are needed, networks between industry and academia are also desirable and important: The statement \textit{Creating networks between research groups and industry will be essential} (S11) gained similar high agreement. 
We need close collaborations of industry and educators. Additionally, industrial needs have to be taken into account in educational efforts, and industrial companies should offer opportunities to educate the future quantum workforce as well.

\subsubsection{Quantum technologies in everyday life}
For the society in general, an increasing relevance of QT is expected (P3), see Tab.~\ref{tab:statem-over}. In this context, there were two related statements: Most participants agreed that \textit{it  will  be  necessary  to  transform  2nd  gen  QTs  from  a research subject to a subject of everyday life}~(S9). 
As well as \textit{it will be necessary to communicate about the transformation} (S10), i.e.~outreach. The second statement gained higher agreement. About twice as many participants rated the outreach item higher than the transformation necessity item than the other way around.
This shows the high importance to make the QT developments visible to society in the future, e.g.~with outreach initiatives.

This becomes even more clear with the statements S13, S14 and S15: Most participants agreed that \textit{2nd gen QTs will contribute to solve everyday problems}~(S13), see Tab.~\ref{tab:statem-over}. 
It is to be hoped that QTs will have a practical use in daily life, in solving everyday problems. Likewise it would be good if they \textit{will contribute to solve social challenges} (S14) -- a statement that found predominantly agreement, see Tab.~\ref{tab:statem-under}. However, for  statement S15, \textit{2nd gen QTs will lead to social inequality}, again a very heterogeneous assessment emerges,  see Tab.~\ref{tab:sat-everydaylife}: 
The highest number of ratings was on rating~3 (rather agree),  though there were similar numbers of votes for all ratings on the disagreement side of the scale.

\begin{table}[hbt]
\caption{\label{tab:sat-everydaylife}  Extract from Tab.~\ref{tab:statem-under}: Rating distribution for Statement S15: \textit{2nd gen QTs will lead to social inequality}.}
\begin{ruledtabular}
\begin{tabular}{rcc|dddddd}
&N&~part~ &1&2&3&4&5&6\\ 
&total&~agree~ & + &\multicolumn{4}{c}{in percentage}&-\\
\hline
S15&57&42&7&11&\cellcolor{black!20}25&19&18&\cellcolor{black!20}21\\ 
\end{tabular}
\end{ruledtabular}
\end{table}

In conclusion, the judgements of the participants on the statements discussed in this paragraph suggest that effects on the broad society are to be expected and outreach activities are needed.

\subsubsection{Profile of a QT curriculum \label{sec:curriculum}}
For the curriculum distribution, Tab~\ref{tab:curriculum} shows the highest agreement is for a nearly equal distribution of educational efforts to all four QT areas. If there should be a stronger focus on one QT area, it should be on quantum computing. 
With respect to communication,  it is notable that there were about as many people voting for more communication ($15\%$ for `$>30$') as for less communication ($21\%$ for `$<20$'). The community is divided on the need for less or more efforts in this area, which shows the dissent and uncertainty in the community. 

However, as for the statements on which of the QTs will become most important, there might be a bias due to the participants' background.  
It is noticeable that the order based on the mean values -- 1st computing, 2nd communication and 3rd sensing -- is the same as for the question \textit{which area will become the most important} and the competence self-assessments. So these data can't be used to justify a strong focus on quantum computing.

\section{Discussion and outlook}\label{sec:discussion}
We collected nearly 200 responses from QT newcomers to experts from  all over Europe. Most of the participants had a scientific background, only about one third had an industrial background, and less than 20\% of the participants assigned themselves to cover the area application/use. This limits our study on the future quantum workforce, people in quantum industry, as we have a strong academic perspective in our results. 

Our research objectives are R1:~\textit{\Rone} and R2:~\textit{\Rtwo}.

With the collected competences (R1), the beta version of the European Competence Framework for Quantum Technologies was compiled~\cite{greinert_nee_gerke_beta_2020, greinert_competence_2021, gerke_vh_greinert_ermittlung_2021}. The collected data comprise three pillars: (a) theoretical background, (b)~practical background and (c) applications (QTs). This structure is visible in the framework. 
Based on what we saw in the main~1 data, there was a pronounced focus on computing in the beta version. This was criticized in the feedback collection in the community. 

\begin{figure}[htb]
\centering
\includegraphics[width=\linewidth]{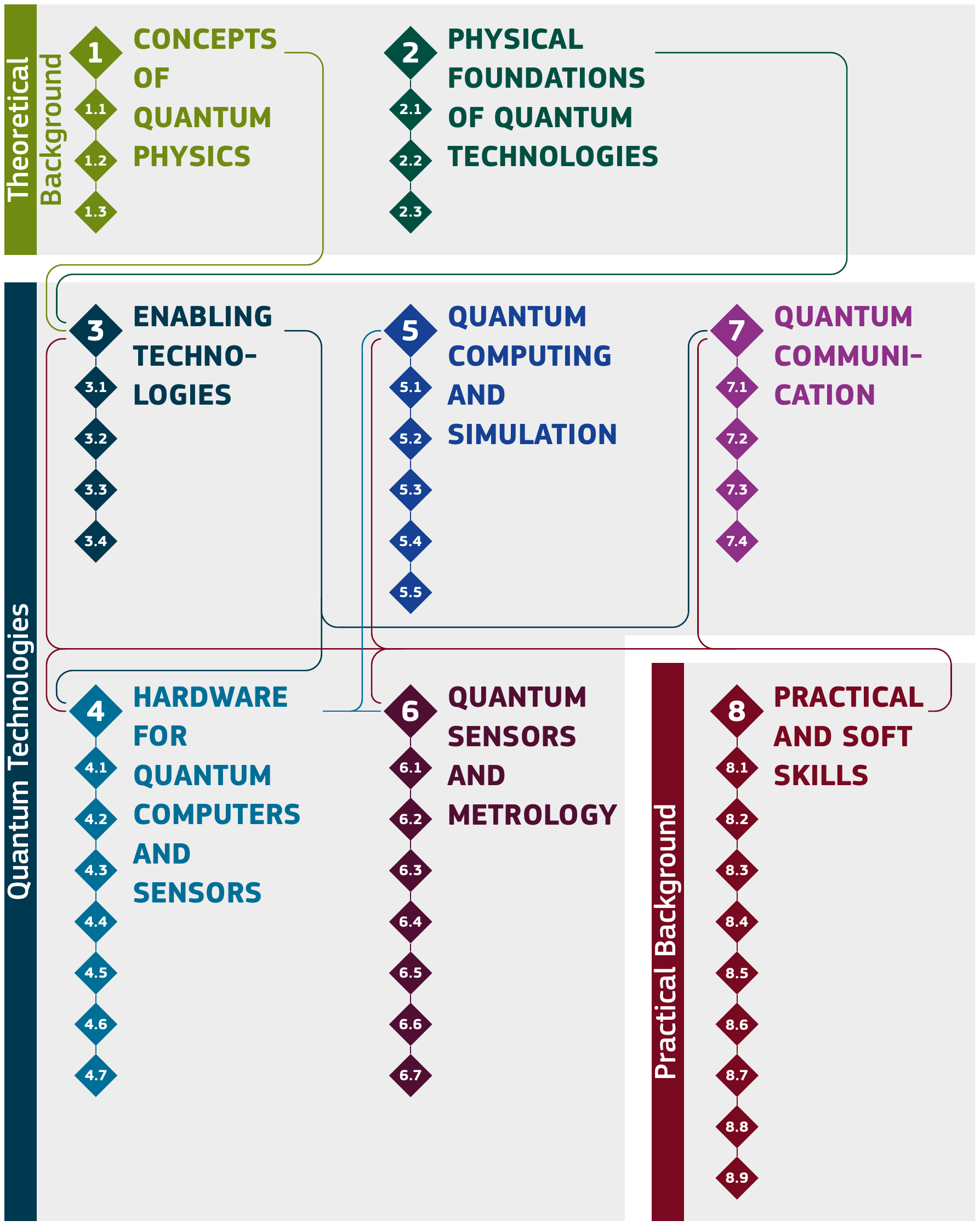}
\caption{Structure of the European Competence Framework for Quantum Technologies version~1.0, see 
\cite{greinert_european_2021}. For reasons of readability, again only the eight main domains explicitly shown. The the anchor points (1.1, 1.2, etc.) for the subdomains give an impression of the overall structure (full-page version with readable subdomains in the appendix, Fig.~\ref{fig:CF-full}.). 
\label{fig:CF}}
\end{figure}

For the detailed revision of the framework domains and the according items, expert interviews were conducted in Spring 2021. As described in the framework's supplemental material~\cite{greinert_competence_2021}, ten interviews in small groups (approx. 3-5 experts per group) were conducted. In the interviews, additional items and suggestions for restructuring or renaming were collected. These led to the version~1.0, published in May 2021. Fig.~\ref{fig:CF} shows the structure of the Competence Framework in version~1.0 after a graphical update in August 2021~\cite{greinert_european_2021}. 

The competences represented in version~1.0 of the Competence Framework are mainly content-specific. However, this is only one dimension of a competence framework. 
At least a second dimension is to be considered, namely the proficiency levels, as has previously been done e.g.~for the  European Framework for the Digital Competence of Educators (DigCompEdu)~\cite{european_commission_joint_research_centre_european_2017}, which was used as a template. 
In version 1.0 they are missing, but were included in the next update (version~2.0 from April 2023, for the latest version see doi \href{https://doi.org/10.5281/zenodo.6834598}{10.5281/zenodo.6834598}). Another aspect are qualification profiles: specific selections of items together with an according proficiency level as examples of which competences a person can reach during educational activities. A beta version from January 2022 of sample qualification profiles is available~\cite{greinert_qualification_2022}.

In the questionnaires, we used the Quantum Flagship's~\cite{quantum_flagship_qteu_2022} four pillars of QTs: computation, simulation, sensors/metrology and communication. A special case is the simulation pillar, as it may be regarded as special case of quantum computing/algorithms. 
This is also how it is shown in the framework version~1.0, where quantum simulation is in domain~5 together with quantum computing (software), and the hardware for quantum computers and sensors has a dedicated domain~(4). This separation between hardware and software is also a topic that could be discussed in further research.

Regarding the predictions (R2), we have seen that the different QT areas will likely evolve at different paces from an expert point of view: While quantum sensing and quantum communication are already important today, computing will probably become more important in the future. However, which one will become the most important QT -- and where the educational programs should focus on -- is not very clear yet, as we have to assume the assessments as biased by the background of the participants. Nevertheless, a tendency that quantum computing will become the most important QT area and should slightly be in the focus in education is recognizable.

This current status is the experts' view for the year 2021. We have tried to figure out what will be relevant in the future in order to plan the current educational activities in such a way that the future workforce will be prepared for the expected demand in the best possible way. Since predictions for the future are always associated with uncertainty, we have to wait and see what developments the next few years will bring and adjust the educational activities accordingly. Especially in the field of QT, we have to assume a great amount of uncertainty, which is also reflected in the sometimes very divided opinions of experts.

However, the data clearly show the need for efforts in educating the future quantum workforce, with special educational programs and networking being needed. In addition, outreach efforts are necessary to avoid what more than 40\% of the participants in main~2 expect: that QTs will lead to social inequality. 
Here QTEdu~\cite{qtedu_csa_qtedueu_2022} with their databases of programs and materials, their community building efforts and pilot projects is a good starting point. For example, a pilot project on outreach reviewed the landscape of quantum education~\cite{seskir_quantum_2022}. Such games can be a gateway to QTs and create initial QT awareness in society.

For the future quantum workforce, special educational programs are needed. One part of the workforce will have a master's degree. Here the EU-founded project DigiQ (Digitally Enhanced European Quantum Technology Master), which emerged from a QTEdu pilot project, will have a coordinating and supporting role. Educational offers ranging from small modules to entire courses will be mapped or developed using the Competence Framework and the Qualification Profiles as the common foundation, while also evaluating them and providing input for updates.
It is also important to develop training and upskilling programs for persons who are already working in the industry and have expertise outside of QT. Such activities will be developed in a coordinated way and made accessible through another project founded by the EU, also based on a QTEdu pilot project: QTIndu (Quantum Technologies courses for Industry). Here the framework will be used for mapping the courses and make them comparable.

Version 1.0 of the framework mainly contains contents. Concrete, measurable competences for different proficiency levels, corresponding learning goals and exemplary examination tasks will be added to the framework in the follow-up Quantum Flagship coordination project QUCATS. With this additions, the framework will be the basis for a certification scheme for training programs and best practice guidelines. This scheme is thought to keep different programs and certificates comparable.

\section*{Acknowledgments}
We thank the QTEdu team for their helpful feedback and input, especially Oxana Mishina, Carrie Weidner and Simon Goorney.

This work has received funding by the European Union’s Horizon 2020 research and innovation programme under grant agreement No 951787. \hfill \raisebox{\dimexpr-\totalheight+\ht\strutbox\relax}{
	\includegraphics[width=1.5cm]{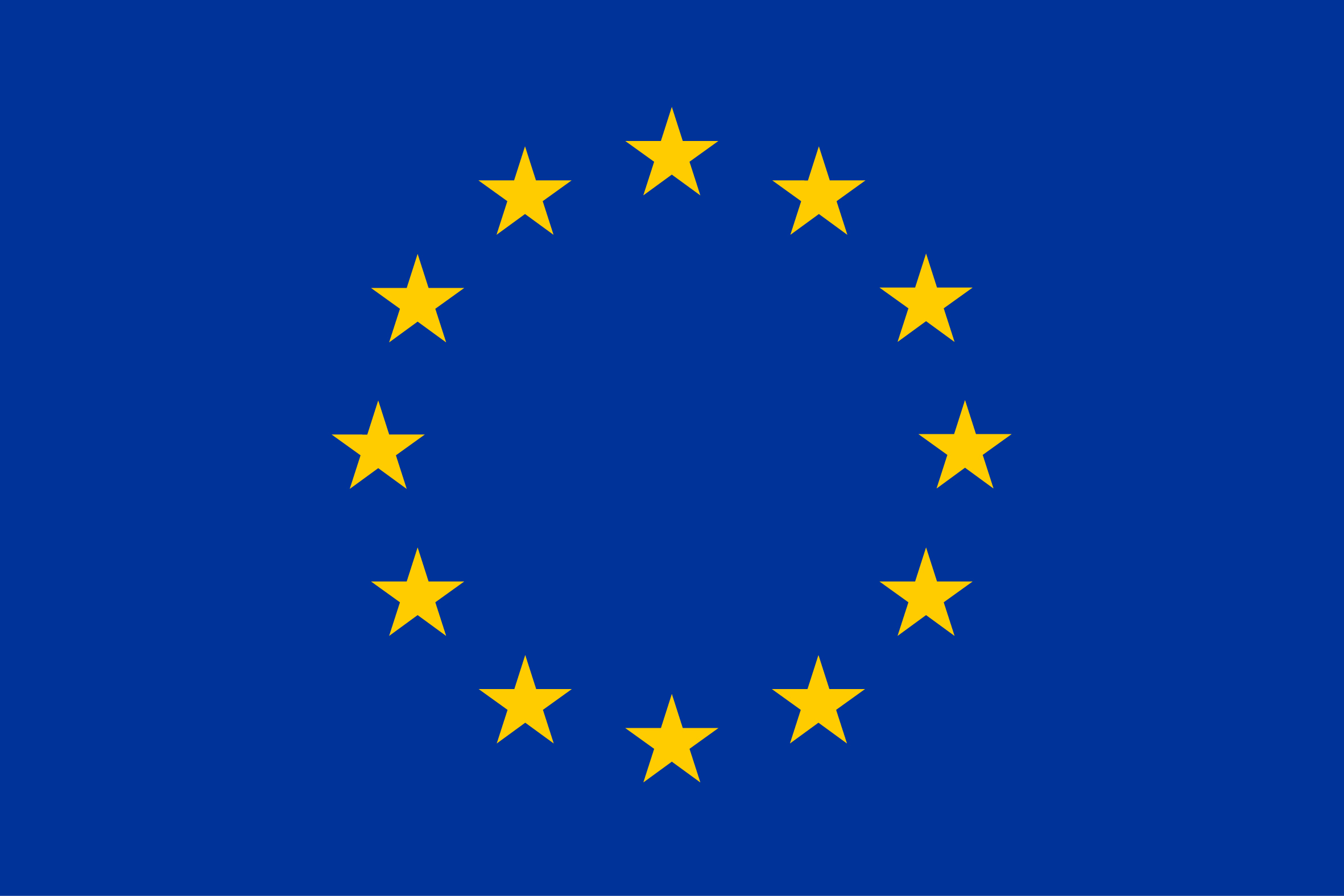}
}
This publication reflects only the views of the authors, the European Commission is not responsible for any use that may be made of the information it contains.

We acknowledge support by the Open Access Publication Funds of Technische Universität Braunschweig. 
~

\section*{Data availability}
Anonymized data from the study is available on request through the Zenodo repository: Requirements for the future Quantum Workforce: Questionnaires and answers from the study, \href{https://doi.org/10.5281/zenodo.6834713}{10.5281/zenodo.6834713}.

\bibliography{references.bib}

\newpage
\appendix

\phantomsection
\section*{Appendix\label{appendix}}
\subsection{Details on the expert panel \label{app:expert}} 
In addition to tables showing the data from rating scale items, we use a bar chart to show the results of a multiple choice question and thus visualize the shift through the rounds, and a map of Europe for the distribution over the countries. 

\begin{figure}[ht]
\centering
\includegraphics[width=0.95\linewidth]{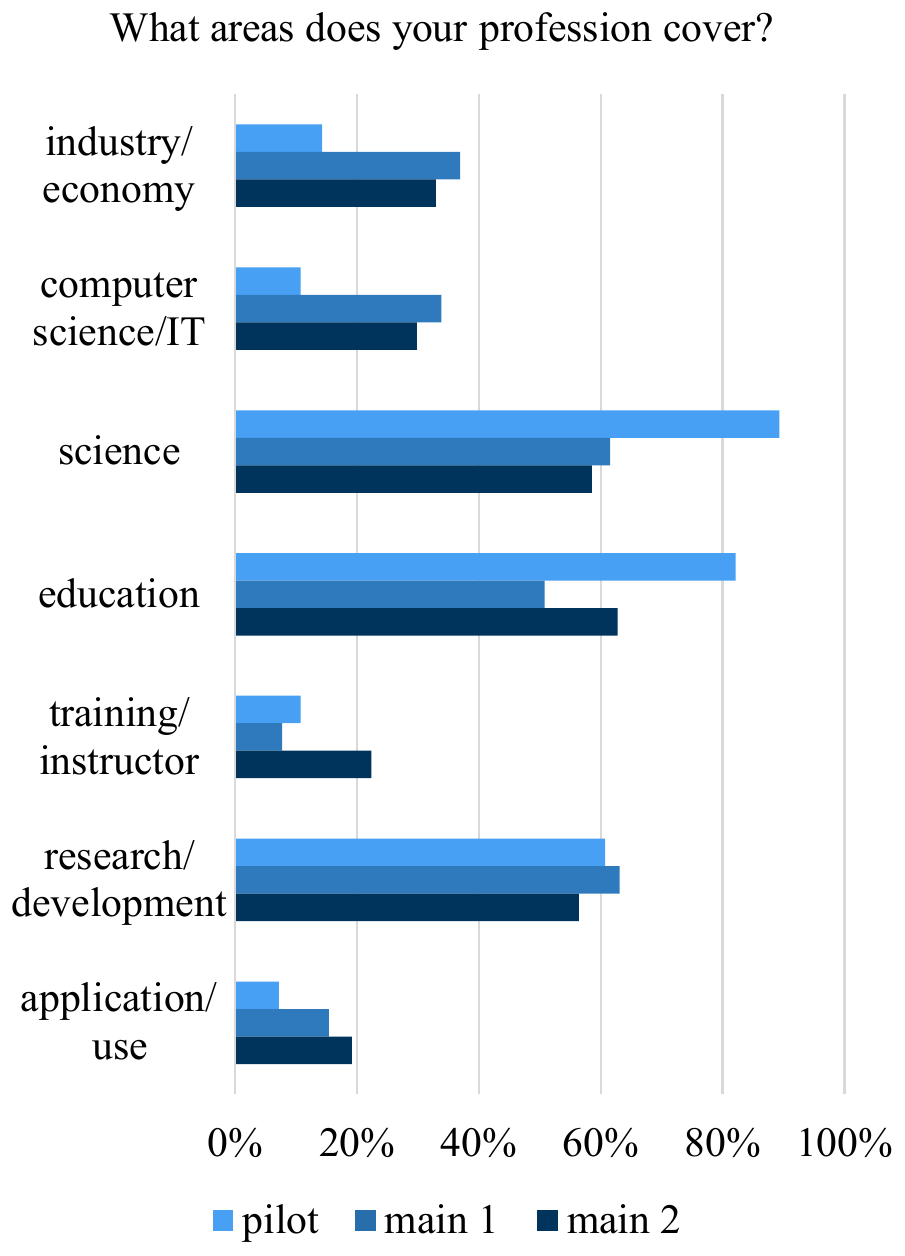}
\caption{Bar chart showing the distribution of the participants on profession areas. In each three-pack the part of the 28 participants that assigned themselves to the area is shown in the upper bar, the bar in the middle belongs to the 65 participants from main~1 and the lower bar belongs to the 94 participants from main~2.}
\label{fig:profback}
\end{figure}

The distribution of the professional background areas is shown in Fig.~\ref{fig:profback}. There is a clear shift between pilot and main~1 and only a subtle shift between the two main rounds.
The main shift between the pilot and the two main rounds is from science and education to industry and computer science.

In Tab.~\ref{tab:owncomp} the participants' competence self-assessment ratings for the three rounds are shown. 
In all rounds, more than two-thirds of the participants indicated a rather to very high self-assessed competence (rating~3, 2 or 1, `part high' in the table). 

\begin{table}[ht]
\caption{\label{tab:owncomp} Participants' competence self-assessment ratings, distribution on a six-level scale from 1 (very high) to 6 (very low) in percent for the three rounds on the right hand side of the vertical line, with the general rating for main~2. In addition, the total number of answers N and the part that rated rather to very high competence (3, 2 or 1) is given.}
\begin{ruledtabular}
\begin{tabular}{l cc|dddddd}
&N&part&1&2&3&4&5&6\\	
&total&high  & + &\multicolumn{4}{c}{in percentage}&-\\
\hline
pilot&28&86&32&43&11&11&4&0\\	main 1&64&69&28&27&14&16&11&5\\	main 2&92&78&20&27&32&12&8&2\\
\end{tabular}
\end{ruledtabular}
\end{table}

\begin{table}[ht]
\caption{\label{tab:howlong} Distribution of the responses for the single-choice question ``How long have you worked in a profession with quantum context?'' with the four time periods $0-3$~years, $3-10$ years, $10-20$ years and more than $20$ years as possible answers, in percentage for all three rounds. In addition, the total number of answers and the part of responses on more than 10 years are given.}
\begin{ruledtabular}
\begin{tabular}{lc c |cccc}
&N&~part~& $> 20$&$10-20$&$3-10$& $0-3$\\
&&~$> 10$ ~&years&years&years&years\\
&& years &\multicolumn{4}{c}{in percentage}\\
\hline
pilot&28&79&36&43&11&11\\	main 1&63&41&19&22&29&30\\	main 2&88&44&19&25&27&28\\
\end{tabular}
\end{ruledtabular}
\end{table}

Tab.~\ref{tab:howlong} shows how long the participants have worked in a profession with quantum context. The relative proportion of experts having more than 10 years of experience nearly halved between the pilot and the two main rounds. Starting with many long-year experts in the pilot, this shifted to more newcomers in main~1 and balanced to a nearly equal distribution across the four surveyed time periods in main~2.

\begin{table}[ht]
\caption{\label{tab:owncompM2} Rating of participants' self-assessed competence in main~2 for the specific areas, sorted by the proportion of participants who rated their own competence rather to very high (rating 3, 2 or 1) with the total number of responses (N) and the percentage of the specific ratings from 1 (very high) to 6 (very low). }
\begin{ruledtabular}
\begin{tabular}{p{0.25\linewidth}@{\hspace{-2pt}}cc|dddddd}
&N&~part~&1&2&3&4&5&6\\
&total&~high~& + &\multicolumn{4}{c}{in percentage}&-\\
\hline
theoretical knowledge&92&77&18&29&29&13&8&2\\
basic/enabling technologies&86&67&12&26&30&12&13&8\\
quantum computing&93&63&9&28&27&14&16&6\\
experimental/ practical skills&90&59&9&24&26&13&14&13\\
quantum communication&89&53&6&20&27&15&22&10\\	
quantum simulation&88&45&10&14&22&22&20&13\\	
quantum sens\-ing/metrology&90&43&10&18&16&20&18&19\\	
\end{tabular}
\end{ruledtabular}
\end{table}

Only in the second main round we asked in detail for the self-reported competence in the QT areas and for theoretical and practical skills. 
This is because in main~1, when we asked for competences for a specific QT subfield, we got a lot of answers from the quantum computing field. So we decided for a more detailed questioning in main~2.  

Tab.~\ref{tab:owncompM2} shows the ratings distribution. It is sorted by the percentage of rather to very high competence ratings, i.e. the part with rating 3, 2 or 1 on the scale from 1 (very high) to 6 (very low). 

\begin{figure*}
\centering
\includegraphics[width=0.81\linewidth]{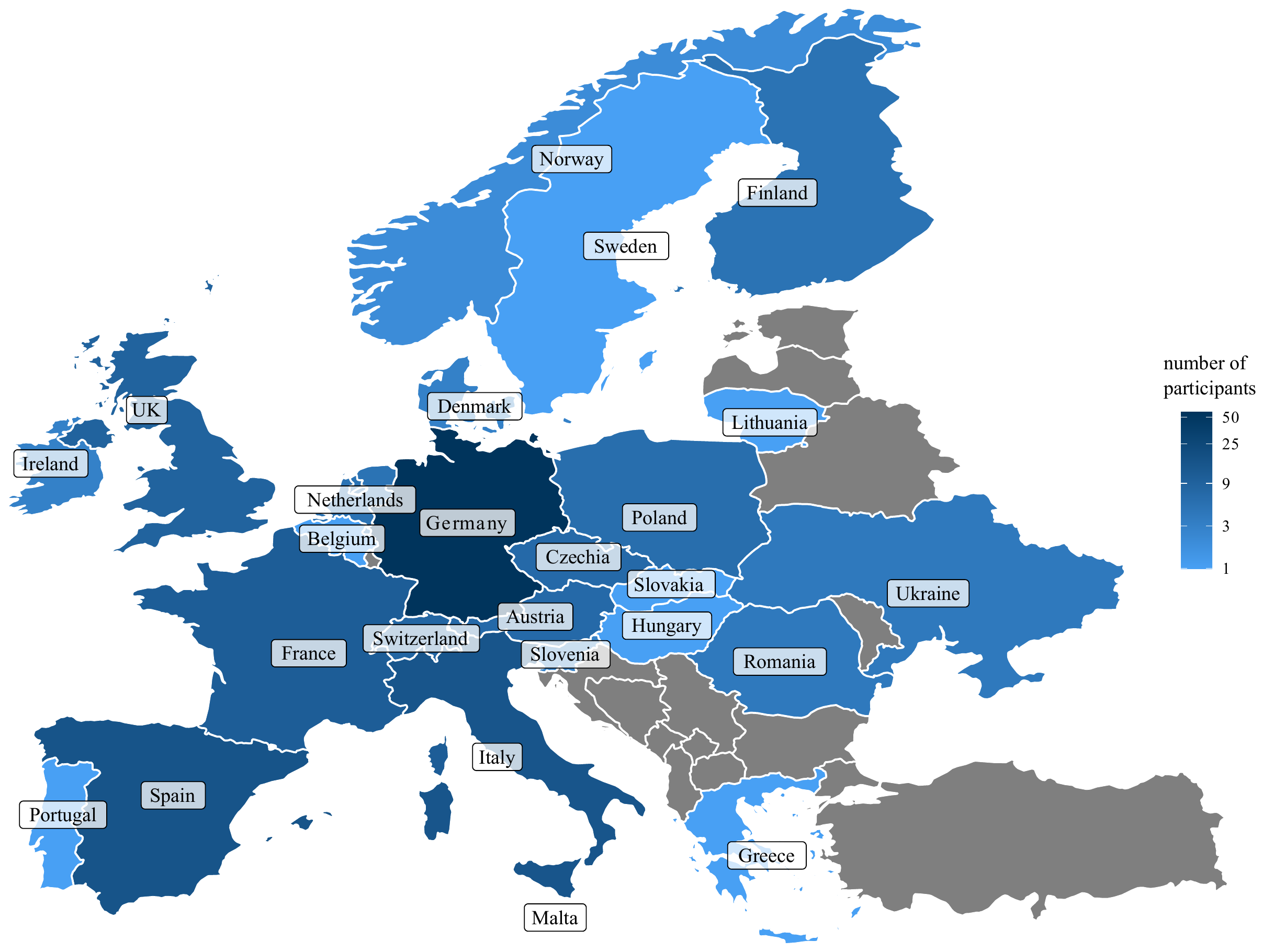}
\caption{Distribution of the 177 participants from Europe that indicated the countries in which they work, together for all study rounds and with the highest number of 57 in Germany.}
\label{fig:countries}
\end{figure*}

The map in Fig.~\ref{fig:countries} shows the distribution of the participants across Europe.
More than 90\% of the participants who indicated a country work in Europe, more than 80\% in EU-countries, and most non-EU experts work in UK and Switzerland. Only a handful are from non-European countries. 
Most participants were from Germany and south-western Europe, but there were also participants from the North, e.g.~from Finland, and the East, e.g.~from Ukraine.

%To map the development of the expert panel, we asked whether they had already participated in the previous rounds. We were able to track 5 participants of main~1 in the pilot dataset based on the provided background data. In main~2, we tracked 12~participants from main~1 and 5 from the pilot, with only one person that participated in all three rounds. However, about twice as many participants indicated that they had participated in a previous round. This leads to estimated 160 different participants.

We see the increasing number of participants as an indicator of the growing interest in the topic, but also of the success of the networking and community building activities which were carried out as part of the QTEdu CSA~\cite{qtedu_csa_qtedueu_2022} and in the Quantum Flagship~\cite{quantum_flagship_qteu_2022}. %In addition, the European Quantum Industry Consortium (QuIC)~\cite{european_quantum_industry_consortium_euroquicorg_2022} distributed invitations to the main rounds to its members. Futhermore, social media channels and the Quantum Flagship~\cite{quantum_flagship_qteu_2022} newsletter were used to announce the study. 
This openness to participants is one reason for the detailed questions on professional backgrounds, as the group of participants was not reduced to a certain pre-selected group.

\begin{table*}[ht]
\subsection{Additional tables for the research objectives\label{app:tabs}}
\vspace{-20pt}
\caption{\label{tab:categoRulesExamples} Category system from main~1 with rules/descriptions of the subcategories and examples. }
\begin{ruledtabular}
\begin{tabular}{p{0.22cm}p{2.23cm}p{5.5cm} p{9.4cm}}
& subcategory&description&examples: ``competence / useful for (level of expertise)''\\
\hline
\multicolumn{4}{l}{\textit{ theoretical background}}\\
&quantum phy\-sics, concepts, phenomena&quantum physical concepts, phenomena, quantum objects, detached from applications (if application references/contexts are mentioned, then double code), incl. qubit, but without gates/algorithms (those belong to CS)&$\cdot$ Knowing phenomenology under the sensing principle (squeezing, entanglement, superposition, quantum interference) (D: deep conceptual and mathematical knowledge to understand potential limits of the device)  \newline
$\cdot$ understaing of physical phenomens in qubits (knowledge of energy levels, Bloch sphere, the ways how EM field interacts with qubits)\\

&classical physics&topics from classical physics (e.g.~semiconductors double code)&$\cdot$ Basic electromagnetism and circuitry (no expertise) \newline
$\cdot$ digital/analog electronics, […]\\

&math &mathematics content, even if it is associated with quantum phenomena (double-code if necessary, e.g.~wave function, Bloch sphere), but the mathematical description is the focus, also notes not to delve too deeply into the mathematics&$\cdot$ linear algebra (U: basic knowledge of vector and matrix operations, tensor products, eigenvalues and matric trace operations, along with a strong intuition (!) of what actions they describe beyond formulas) \newline 
$\cdot$ theoretical description of quantum mechanical systems (states, density matrices, Hilbert-space) (D: deep knowledge essential)\\

&(quantum) computer science&aspects of (quantum) computer science and new approaches from this field (programming skills, gates, algorithms, artificial intelligence/machine learning)&$\cdot$ Understanding of information science and artificial intelligence (D: advanced knowledge of classical and quantum bits and artificial intelligence methods (deep learning, genetic algorithms, etc.)) \newline
$\cdot$ software development in general (e.g., Python, Assembler) (some experience required but not on a very high level (for quantum application developers))\\

\multicolumn{4}{l}{\textit{practical background}}\\
&experimental/ practical skills, phy\-si\-cal/tech\-ni\-cal realisation&practical competences from the field of physics, technical implementation: what is realizable/implementable (preparation of states, possibilities, technical limits, current status)&$\cdot$ practical experience in the generation of photons and their quantum states.   \newline
$\cdot$ understand the state of the art, what can and cannot be done in experimental physics and what the barriers are and why.\\
&engineering/in\-dustrialisation &practical creation aspects (build, design, implementation, manufacturing)&$\cdot$ knowing how to build things that work outside of the lab (passion for engineering and having a functional device at the end of the day -- medium)  \newline
$\cdot$ Atomic precise manufacturing (Master level)\\
&soft skills, social aspects&communication skills, networking, philosophical aspects&$\cdot$ Capability of explaining to policy makers and business community representatives the relevance of quantum technologies to boost economic and social growth.  \newline
$\cdot$ know the programs and projects underway and the main actors of the quantum ecosystem. (U: medium knowledge to know where to find sources of information.)\\

\multicolumn{4}{l}{\textit{application (`useful for' part)}}\\
&engineering applications, production of QTs in general\vspace{4pt}&`useful for' (even if the actual `useful for' appears in the competence or expertise level) applications that cannot be assigned to a specific  main area&$\cdot$ design, control and stabilization of hardware \newline 
$\cdot$ developing new components like single photon detectors \newline
$\cdot$ Understanding costs and difficulties for achieving for the operation of the devices. \\
&application in q. computing&everything `useful for' (see above) with reference to quantum computing&$\cdot$ understanding of quantum-hardware operating parameters (gate operations, connectivity, fidelity, gate-speed, ...) for multiple types of qubit-architectures (superconducting, trapped ions, photonic,~...) (D: quantum-developers need to be fluent in this field) \newline
$\cdot$ basic understanding of quantum computing toolboxes, gates, measurement (U: basic understanding of the ingredients that must be implemented in actual technology.)\\
&application in q. communication, sensing or simulation&everything `useful for' (see above) with reference to quantum communication, sensor technology and simulation as concrete application&$\cdot$ understanding of quantum communication protocols (U: deeper basic knowledge of quantum communication protocols and routing algorithms. Good overall understanding of network architecture. No specific knowledge of physical implementation of hybrid classical/quantum internet. D: deeper understanding of physical hardware needed to implement hybrid classical quantum internet.) \newline
$\cdot$ Implementations of quantum sensors for dedicated applications \newline
$\cdot$ understanding the potential of the quantum simulation platform \\
\multicolumn{2}{l}{\textit{other}}&everything else&\\
\end{tabular}
\end{ruledtabular}
\end{table*}

\newcommand{\Mark}{\cellcolor{black!20}}
\begin{table*}
\caption{Statements with their identifiers that gained more than 80\% agreement in main~2, with the number of answers~(N) and sorted decreasing by the part that agreed to the statement, and with details of the rating on the scale from 1~(total agreement) to 6~(total disagreement) in percentage. \label{tab:statem-over}}
\begin{ruledtabular}
\begin{tabular}{r@{\hspace{-0.2cm}}p{6.5cm}cc|dddddd}
&&N&~part~ &1&2&3&4&5&6\\ 
&Statement&total&~agree~ & + &\multicolumn{4}{c}{in percentage}&-\\
\hline
P1&The relevance of QTs for industry will increase significantly in the near future.&67&99&\Mark{}52&34&12&1&0&0\\
S12&Special educational programs fitted to arising needs are necessary.&64&98&\Mark{}53&31&14&0&2&0\\
P2&QTs are already very important in science, but even here they will become more important in the next few years.&67&97&\Mark{}54&30&13&3&0&0\\
S4&2nd gen QTs will enable further steps in fundamental research.&64&97&\Mark{}47&41&9&2&2&0\\
S11&Creating networks between research groups and industry will be essential.&63&95&\Mark{}52&29&14&5&0&0\\
S8&Decoherence is one of the most central challenges to be addressed in the realisation of 2nd gen QTs.&57&93&\Mark{}44&42&7&5&2&0\\
M10&Quantum communication will become very important because of cryptography/security and use in secure communication in banking, military, politics, etc.&62&92&31&\Mark{}39&23&3&3&2\\
S2&The technological change of paradigm, i.e. the extension of current technologies to hybrid systems, is a really important aspect of 2nd gen QTs.&55&91&20&\Mark{}49&22&7&2&0\\
M9&Quantum sensors/metrology will become very important through use in timing/navigation, observation and autonomous devices/AI.&58&90&22&\Mark{}45&22&7&2&2\\
S10&It will be necessary to communicate about the transformation of 2nd gen QTs from a research subject to a subject of everyday life (outreach).&63&89&33&\Mark{}40&16&10&0&2\\
P3&The relevance of QTs for society will increase significantly in the near future.&68&88&25&\Mark{}40&24&9&3&0\\
S7&The interaction and integration of classical and quantum systems will be in focus of 2nd gen QTs.&58&88&22&\Mark{}45&21&7&5&0\\
M12&Enabling/basic technologies will be the first to become really important in industry, as ``the industrial impact of QT can only been realized when quantum engineering, integration, miniaturization and scaling is realized'', so thier role is ``moving QT from the lab into society, making it aviable at reasonable cost''.&62&87&29&\Mark{}39&19&10&3&0\\
S9&It will be necessary to transform 2nd gen QTs from a research subject to a subject of everyday life.&63&87&25&\Mark{}43&19&5&3&5\\
M7&Quantum simulation will have ``enormous long-term value for chemistry, pharmacy, material science, etc.''&60&85&20&\Mark{}40&25&12&3&0\\
M8&Quantum sensors/metrology will become very important through use in medicine (e.g.~imaging).&59&83&19&\Mark{}41&24&15&2&0\\
M6&Quantum computation has the ``highest gain potential'' of all QT areas. In the long-term, it ``will have more impact and will really be disruptive.''&62&82&31&\Mark{}35&16&6&11&0\\
S13&2nd gen QTs will contribute to solve everyday problems.&62&81&18&\Mark{}35&27&16&0&3\\
\end{tabular}
\end{ruledtabular}
\end{table*}

\begin{table*}
\caption{Continuation of Tab.~\ref{tab:statem-over} with the statements that gained less than 80\% agreement, again sorted by decreasing agreement and with details of the rating on a scale from 1 (total agreement) to 6 (total disagreement) in percentage.  \label{tab:statem-under}}
\begin{ruledtabular}
\begin{tabular}{r@{\hspace{-0.2cm}}p{6.5cm}cc|dddddd}
&&N&~part~ &1&2&3&4&5&6\\ 
&Statement&total&~agree~ & + &\multicolumn{4}{c}{in percentage}&-\\
\hline
M11&Quantum communication will become very important in the context of the quantum internet.&62&77&16&\Mark{}40&21&13&8&2\\
M3&In the long term, quantum computing will become the most important QT area.&66&76&23&\Mark{}38&15&14&6&5\\
M5&In the long term, quantum communication will become the most important QT area.&65&75&17&26&\Mark{}32&12&11&2\\
S1&Quantum chemistry will be the most important subfield of 2nd gen QTs.&61&74&11&28&\Mark{}34&18&5&3\\
M4&In the long term, quantum sensing/metrology will become the most important QT area.&64&64&11&25&\Mark{}28&19&16&2\\
Q1&``For me it is a question of maturity and opportunity window... all the disciplines will be important in the short term... those more matured and more deployed will lose their `importance' because they would have been absorbed, accepted and assimilated in the long term.... other will continue in the top in the long term due to their inmaturity or potential of evolution still for develop...''&53&62&11&25&\Mark{}26&21&13&4\\
S14&2nd gen QTs will contribute to solve social challenges.&61&62&13&21&\Mark{}28&25&8&5\\
Q2&``In my opinion quantum communication including quantum internet will remain a merely academically interesting field of technology assuming that the only application which will be found for it is quantum-secure communication. The reason for this is that post-quantum-crypto systems (which are quantum-safe but classical alternatives to our existing crypto systems) will provide the solution for the risk which quantum poses to existing crypto-systems. Therefore, unless you have some national security type communication, quantum-key-distribution will always remain an unnecessarily expensive alternative to PQC systems. The other QT will in my opinion in the mid/long-term provide important contributions to business and society''&61&57&7&21&\Mark{}30&18&21&3\\
S5&It is more important to push 1st gen QTs to make 2nd gen emerge on this basis than pushing 2nd gen directly.&50&50&8&18&\Mark{}24&18&\Mark{}26&6\\
S6&In practice the emergence of 2nd gen from 1st gen is not essential.&48&48&6&\Mark{}25&17&13&\Mark{}25&15\\
M1&In the near future, quantum computing will be less important than the other QT areas.&67&43&7&\Mark{}25&10&15&\Mark{}24&18\\
S15&2nd gen QTs will lead to social inequality.&57&42&7&11&\Mark{}25&19&18&\Mark{}21\\
M2&In the near future, quantum simulation will be less important than the other QT areas.&63&41&3&14&\Mark{}24&19&\Mark{}22&17\\
S3&Fundamental research becomes less important within 2nd gen QTs.&65&22&3&6&12&23&23&\Mark{}32\\
\end{tabular}
\end{ruledtabular}
\end{table*}

\begin{figure*}[ht]
\subsection{Full overview pages of the Competence Framework\label{app:CFoverv}}
\centering
\includegraphics[width=1.2\linewidth, angle=90]{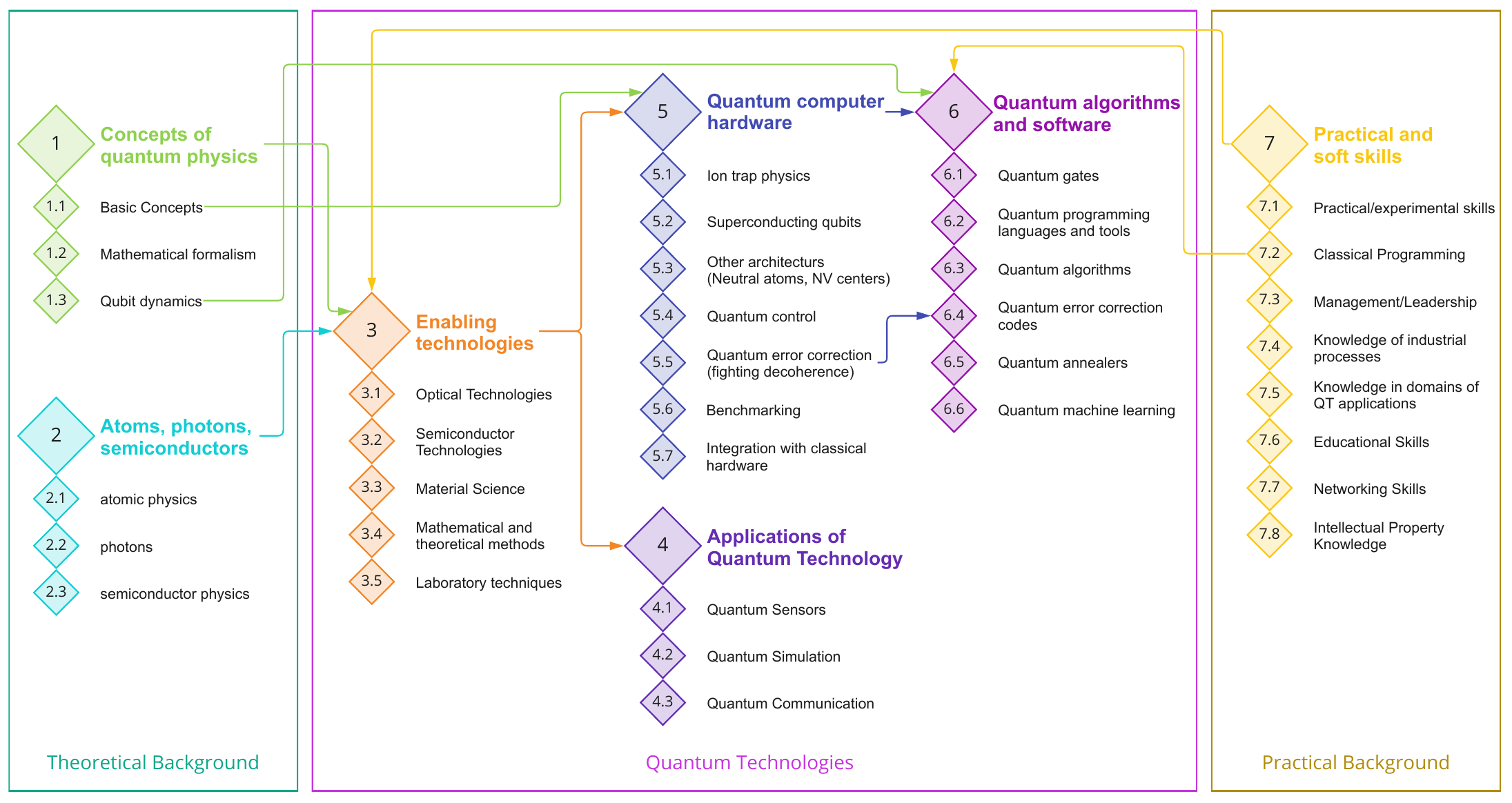}
\caption{Overview of the beta version of the European Competence Framework for Quantum Technologies~\cite{greinert_nee_gerke_beta_2020} from December 2020. \label{fig:CF-beta-full}}
\end{figure*}

\begin{figure*}
\centering
\includegraphics[width=0.97\linewidth]{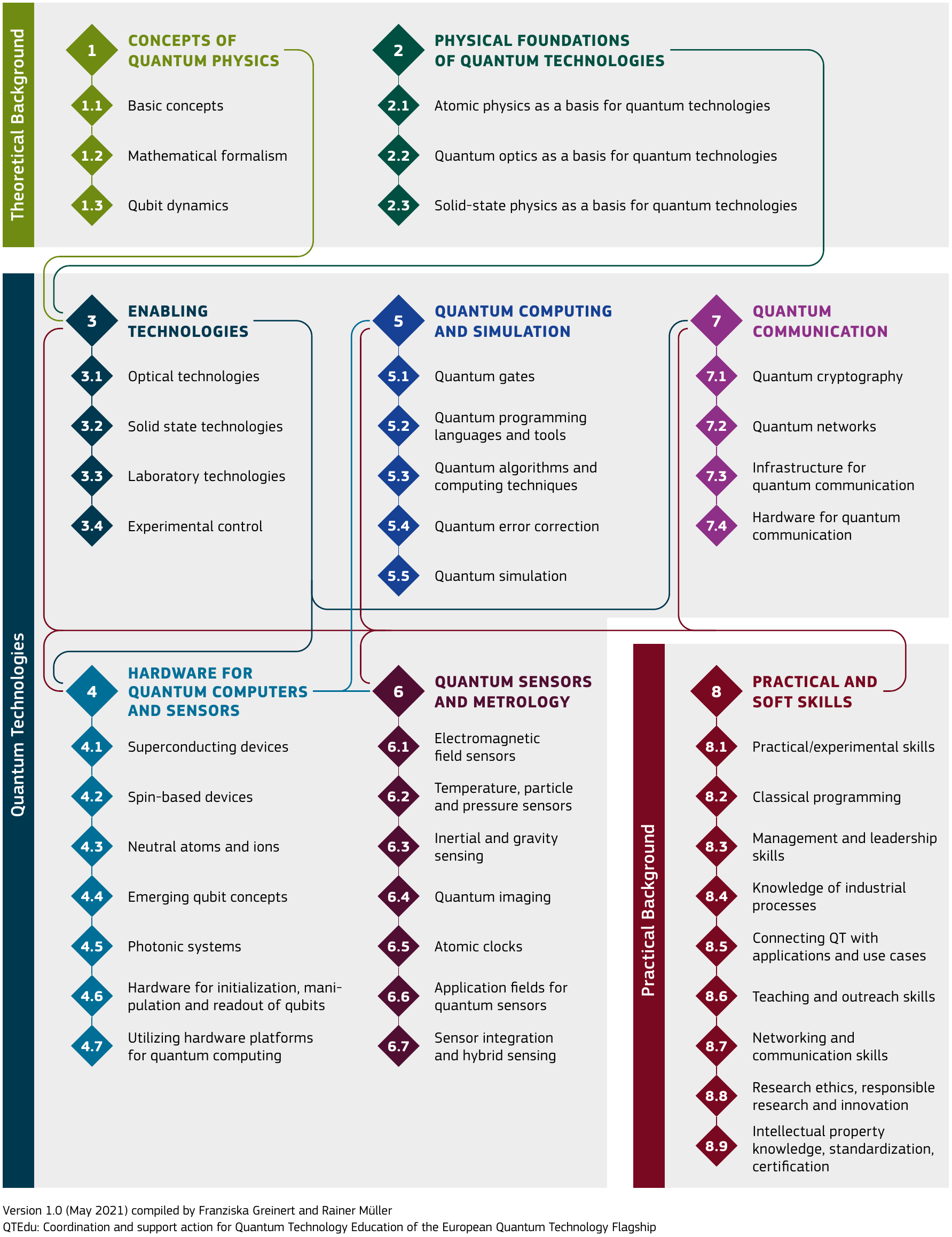}
\caption{Overview of the European Competence Framework for Quantum Technologies version~1.0 
\cite{greinert_european_2021}. 
\label{fig:CF-full}}
\end{figure*}

\end{document}